\documentclass[english,aps,prd,superscriptaddress,twocolumn]{revtex4-1}
\usepackage[T1]{fontenc}
\usepackage[latin9]{inputenc}
\usepackage{amsmath}
\usepackage{graphicx}
\usepackage{xcolor}

\makeatletter
\usepackage{slashed}
\usepackage{hyperref}
\usepackage{MnSymbol}
\usepackage{bbm}

\makeatother

\usepackage{babel}
\begin{document}

\title{Hamiltonian approach to the torsional anomalies and its dimensional
ladder}

\date{\today}

\author{Ze-Min Huang}
\email{zeminh2@illinois.edu}

\affiliation{Department of Physics, University of Illinois at Urbana-Champaign,
1110 West Green Street, Urbana, Illinois 61801, USA}

\author{Bo Han}

\affiliation{Theory of Condensed Matter Group, Cavendish Laboratory, University of Cambridge, 
J.~J.~Thomson Avenue, Cambridge CB3 0HE, 
United Kingdom}

\affiliation{Department of Physics, University of Illinois at Urbana-Champaign,
1110 West Green Street, Urbana, Illinois 61801, USA}

\author{Michael Stone}

\affiliation{Department of Physics, University of Illinois at Urbana-Champaign,
1110 West Green Street, Urbana, Illinois 61801, USA}

\begin{abstract}
Torsion can cause various anomalies in various dimensions, including
the $\left(3+1\right)$-dimensional $[(3+1)D]$ Nieh-Yan anomaly, the $\left(2+1\right)$D
Hughes-Leigh-Fradkin (HLF) parity anomaly, and the $\left(3+1\right)$D,
$\left(1+1\right)$D chiral energy-momentum anomaly. We study these
anomalies from the Hamiltonian approach. We derive the $\left(1+1\right)$D
chiral energy-momentum anomaly from the single-body Hamiltonian. We
then show how other torsional anomalies can be related to the $\left(1+1\right)$D
chiral energy-momentum anomaly in a straightforward way. Finally, 
the Nieh-Yan anomaly and the $\left(3+1\right)$D chiral energy-momentum
anomaly are obtained from the parity anomaly and the HLF effective action, respectively. Hence, we have constructed the dimensional ladder for the torsional
anomalies from the single-body Hamiltonian picture.
\end{abstract}
\maketitle

\section{Introduction}

Many topological phases of matter can be understood from quantum anomalies
\citep{adler1969pr, bell1969nca, ryu2012prb,witten2016rmp, zyuzin2012prb}. A classic example is the integer
quantum Hall effect, where the Hall current is captured by the $\left(2+1\right)$-dimensional
parity anomaly in the bulk and thus the $\left(1+1\right)$-dimensional gauge
anomaly at the edges \citep{stone1992worldscientific}. Both the electromagnetic
anomalies and the gravitational anomalies have been extensively used
to characterize various topological phases of matter \citep{landsteiner2011prl,zyuzin2012prb,son2013prb,furusaki2013crp,gorbar2014prb, li2016naturephy,li2016naturecom,huang2017prb,gooth2017nature,stone2018prd}.
Especially, the dimensional ladder for these anomalies has been known
for a long time \citep{nakahara2003crc}. Namely, the chiral anomaly,
the parity anomaly and the gauge anomaly in different dimensions relate
to each other closely, which has been used to classify topological
phases of matter in different dimensions \citep{qi2008prb,ryu2012prb}.

Torsion can cause quantum anomalies as well, but they are less studied
compared to the electromagnetic anomalies and the gravitational anomalies.
However, torsion is ubiquitous in condensed matter. The torsion tensor
can be used as a probe for thermal transport \citep{luttinger1964pr,tatara2015prl,bradlyn2015prb}
and also to describe dislocations \citep{katanaev1992aop,hughes2011prl,hughes2013prd,parrikar2014prd}.
In addition, torsion can also emerge from a order parameters of
Fermi superfluid or topological superconductors \citep{volovik2003oxford,golan2018prb,Nissinen2019arxivNY}.
Hence, torsional effects in topological phases of matter are attracting
more and more attention and they can lead to various novel phenomena,
including the anomalous thermal Hall effect in Weyl semimetals \citep{gorbar2017prb, huang2019arxiv},
the chiral torsional magnetic effect \citep{sumiyoshi2016prl} and
other viscoelastic responses \citep{hughes2011prl,parrikar2014prd,sun2014epl,shapourian2015prb,you2016prb,palumbo2016aop,huang2018prb,khaidukov2018jetp,nissinen2019prr}.
Due to the close relation between topological phases of matter and
quantum anomalies, it is thus natural to ask what torsional anomalies
 we have and whether they form a dimensional ladder.

Various torsional anomalies have been obtained in the context of high-energy
physics \citep{nieh1982aop,nieh1982jmp,obukhov1997fop,chandia1997prd,chandia1998prd,peeters1999jhep,hughes2013prd,parrikar2014prd},
including the Nieh-Yan chiral anomaly and the chiral energy-momentum
anomaly (or the chiral diffeomorphism anomaly) in four dimensions
\citep{nieh1982jmp,nieh1982aop,chandia1997prd,obukhov1997fop,chandia1998prd,peeters1999jhep,parrikar2014prd},
the Hughes-Leigh-Fradkin (HLF) parity anomaly \citep{FootN1}  in three dimensions
\citep{hughes2011prl,hughes2013prd} and the chiral energy-momentum
anomaly in two dimensions \citep{hughes2013prd}. In the same spirits
as Ref. \citep{alvarez1985aop}, these anomalies can be derived from
the Atiyah-Patodi-Singer index theorem or the chiral anomaly \citep{parrikar2014prd}.
However, the torsional anomalies listed here are all divergent and
thus depend on the ultra-violet physics. Hence, an alternative approach
based on the Hamiltonian is highly desired, which enables us to appreciate
the physics behind these divergences because the Hamiltonian approach is more familiar to
condensed matter physicists.

In this paper, we derive various torsional anomalies based on
the Hamiltonian approach. First, the $\left(1+1\right)$-dimensional chiral
energy-momentum anomaly is derived from the single-body Hamiltonian.
Then, all other torsional anomalies can be straightforwardly obtained
from this anomaly, including the $\left(2+1\right)$-dimensional HLF effective
action, the $\left(3+1\right)$-dimensional Nieh-Yan anomaly and the $\left(3+1\right)$-dimensional
chiral energy-momentum anomaly. Especially, we show that the quadratic
cutoff in the $\left(1+1\right)$-dimensional chiral energy-momentum anomaly
is the energy density, so the cutoff measures the depth of the vacuum.
We also reveal the close relation between the $\left(3+1\right)$-dimensional
torsional anomalies and the $\left(2+1\right)$-dimensional parity-odd effective
action. Namely, assuming translational invariance along one direction,
the corresponding momentum is a good quantum number, so it acts like
a mass term. The response currents from the $\left(2+1\right)$-dimensional
effective action are then used to derive the corresponding $\left(3+1\right)$-dimensional
anomaly currents at a given momentum. The corresponding $\left(3+1\right)$-dimensional
anomaly current can thus be obtained by summing over all the momentum.
Finally, motivated by the recently predicted thermal Nieh-Yan anomaly
\citep{nissinen2019arxivWM,huang2019arxiv,nissinen2019arxivSF}, we
conjecture that there are thermal corrections to the torsional anomalies
in the dimensional ladder obtained above, where the dimensionful coefficients
are replaced by the temperature. Although we have focused on the $4$--$3$--$2$ dimensional ladder, our results can be easily generalized to $(2n+2)$--$(2n+1)$--$(2n)$ dimensional ladder, which is discussed in the two appendixes.

\begin{figure}
\includegraphics[scale=0.25]{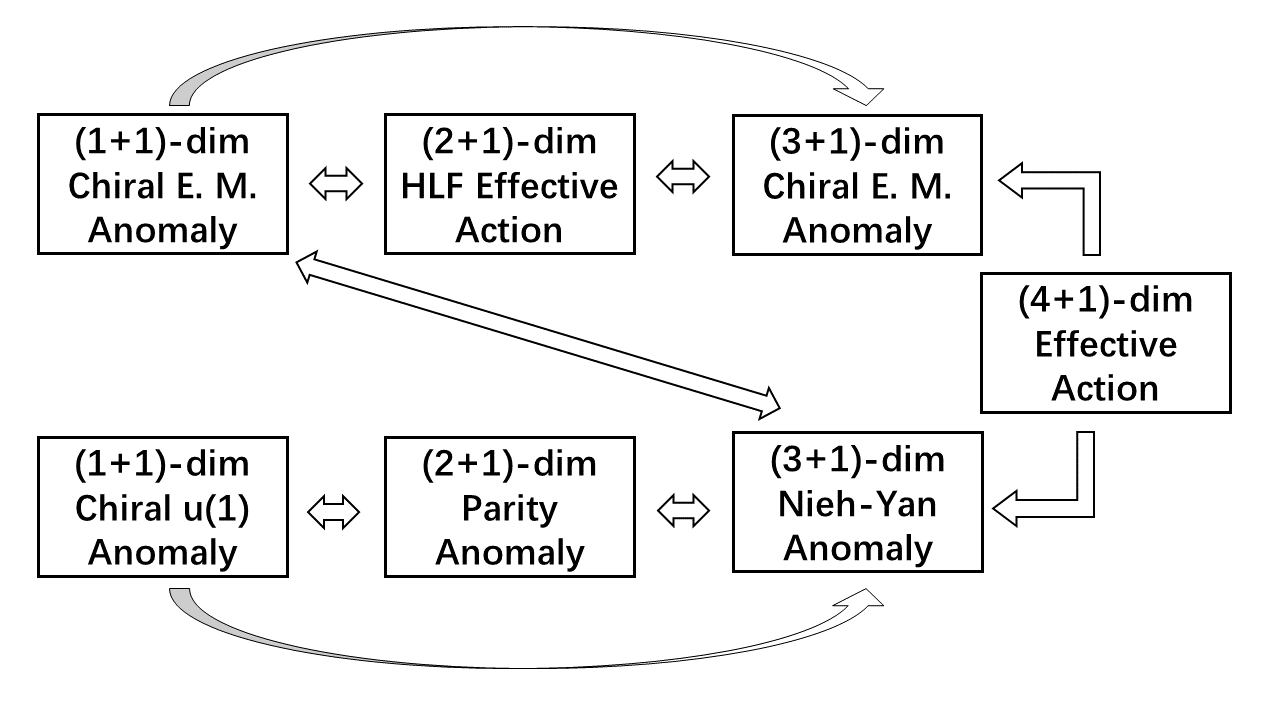}\caption{Dimensional ladder for the torsional anomalies\label{fig:Dimensional_ladder}.
``Chiral E. M. Anomaly'' stands for the chiral energy-momentum anomaly.
``HLF effective action'' stands for the Hughes-Leigh-Fradkin effective
action obtained in Ref. \citep{hughes2011prl}. The boxes in the first
line are for the anomalies related to the energy-momentum tensor (the corresponding derivation based on the Hamiltonian approach is given in Sec. \ref{sec:ChEM_anomaly}), while boxes
in the bottom are for the anomalies in the chiral U(1) current (the corresponding derivation is presented in Sec. \ref{sec:Ch_current_anomaly}). }
\end{figure}

The rest of this paper is organized as follow. In Sec. \ref{sec:summary_torsionalanomalies},
we give a brief summary of the torsional anomalies. In Sec. \ref{sec:ChEM_anomaly},
we construct the first part of the dimensional ladder shown in Fig.
\ref{fig:Dimensional_ladder}, i.e., the $\left(1+1\right)$-dimensional chiral
energy-momentum anomaly, the $\left(2+1\right)$-dimensional HLF effective
action, and the $\left(3+1\right)$-dimensional chiral energy-momentum anomaly.
In Sec. \ref{sec:Ch_current_anomaly}, we construct the second part
of the dimensional ladder for the chiral current, i.e., $\left(1+1\right)$-dimensional
chiral $U(1)$ anomaly, $\left(2+1\right)$-dimensional parity anomaly, and $\left(3+1\right)$-dimensional
Nieh-Yan anomaly. We summarize the main results of this paper in Sec.
\ref{sec:summary_dis}.

\section{Torsional anomalies \label{sec:summary_torsionalanomalies}}

It is known that torsion can induce various anomalies, including the
Nieh-Yan anomaly in four dimensions \citep{nieh1982aop,nieh1982jmp,chandia1997prd},
the chiral energy-momentum anomaly in two and four dimensions \citep{hughes2013prd,parrikar2014prd}
and the HLF parity anomaly in three dimensions.
Interestingly, all these anomalies contain a divergent coefficient
and thus depend on the ultraviolet cutoff. To be more concrete,
the Nieh-Yan anomaly is (for the derivation using Fujikawa's method,
please refer to Appendix \ref{sec:Nie_Yan_fujikawa})

\begin{equation}
\frac{1}{\sqrt{\left|g\right|}}\partial_{\mu}\sqrt{\left|g\right|}j^{5\mu}=\frac{\Lambda^{2}}{32\pi^{2}}\frac{\epsilon^{\mu\nu\rho\sigma}}{\sqrt{\left|g\right|}}\left({T^{a}}_{\mu\nu}{T^{b}}_{\rho\sigma}\eta_{ab}-2e_{\mu}^{*a}e_{\nu}^{*b}\Omega_{ab\rho\sigma}\right),\label{eq:Nieh_Yan}
\end{equation}
where $\Lambda$ is the cutoff, $e_{\mu}^{*a}$ is the vielbein and
$g_{\mu\nu}$ is the metric with $g\equiv\det g_{\mu\nu}$. $a,\thinspace\mu=0,\thinspace1,\thinspace2,\thinspace3$
are the coordinates indices: $a$ is called the Lorentz index and
it labels the locally flat coordinates with basis $\left\{ e_{a}\right\} $,
while $\mu$ is called the Einstein index and it is used for the basis $\left\{ \partial_{\mu}\right\} $.
$j^{5\mu}$ is the axial current. ${T^{a}}_{\mu\nu}$ is the torsion
tensor, i.e., ${T^{a}}=\frac{1}{2}{T^{a}}_{\mu\nu}dx^{\mu}\wedge dx^{\nu}$
and ${T^{a}}=de^{*a}+{\omega^{a}}_{b}\wedge e^{*b}$, where ${\omega^{a}}_{b}={\omega^{a}}_{b\mu}dx^{\mu}$
is the spin connection. $\Omega_{ab\mu\nu}$ is the curvature tensor,
i.e., ${\Omega}_{ab}=\frac{1}{2}{\Omega}_{ab\mu\nu}dx^{\mu}\wedge dx^{\nu}$
and $\Omega_{ab}=d\omega_{ab}+{\omega_{a}}^{c}\wedge\omega_{cb}$.

As for the massless Dirac fermions in $\left(1+1\right)$-dimensional space-time,
we have (for details, refer to Appendix \ref{sec:chiral_energy_momentum_anomaly})
\begin{eqnarray}
 &  & \nabla_{\mu}\tau_{\left(c\right)a}^{5\mu}+{T^{\rho}}_{\mu\rho}\tau_{\left(c\right)a}^{5\mu}-e_{a}^{\mu}{T^{d}}_{\mu\nu}\tau_{\left(c\right)d}^{5\nu}\nonumber \\
 & = & \left(\frac{\Lambda^{2}}{4\pi}\right)\eta_{ab}\frac{\epsilon^{\mu\nu}}{\sqrt{\left|g\right|}}{T^{b}}_{\mu\nu}+\dots,\label{eq:chiral_em_anomaly_2}
\end{eqnarray}
where $\dots$ means that we have kept only the leading divergent terms.
$\tau_{\left(c\right)a}^{\mu}$ and $\tau_{\left(c\right)a}^{5\mu}$
are the canonical energy-momentum tensor and the chiral canonical
energy-momentum tensor, respectively. $\eta_{ab}=\text{diag}\left(1,\thinspace-1\right)$
is the metric. Terms in the first line of Eq. (\ref{eq:chiral_em_anomaly_2}) can be derived from the chiral covariant translational symmetry. That is,  the right-handed Weyl fermions transform as $\psi_{R}\rightarrow \xi^{\mu}D_{\mu}\psi_{R}$, and $\psi_L\rightarrow -\xi^{\mu}D_{\mu}\psi_{L}$ for left-handed Weyl fermions (for details, 
refer to Appendix \ref{sec:chiral_energy_momentum_anomaly}), where $D_\mu$ is the Lorentz covariant derivative and it acts on the Lorentz indices. So these
terms appear even at the classical limit. Compared with the U(1) case, fermions coupled to the electromagnetic field have a classical equation of motion for the energy-momentum tensor as $\nabla_{\mu}\tau^{\mu\nu}=F^{\nu\mu}j_{\mu}$, where $F^{\nu \mu}$ is the field strength tensor of the electromagnetism. In both cases, we treat the electromagnetic field or the torsion field as the external background field. The terms in
the second line are the anomalies from the quantum fluctuations. In the same spirit as the Callan-Harvey mechanism \citep{callan1985npb},
this term is expected to originate from the current inflow in the
bulk, which will be explored in the next section. This chiral energy-momentum
anomaly appears in four dimensions as well. Especially, for the massless
Dirac fermions, 
\begin{eqnarray}
 &  & \nabla_{\mu}\tau_{\left(c\right)a}^{5\mu}+{T^{\rho}}_{\mu\rho}\tau_{\left(c\right)a}^{5\mu}-e_{a}^{\nu}\left({T^{d}}_{\mu\nu}\tau_{\left(c\right)d}^{5\mu}-\Omega_{cd,\thinspace\mu\nu}S^{\mu cd}\right)\nonumber \\
 & = & \frac{\Lambda^{2}}{16\pi^{2}}\eta_{ab}\frac{\epsilon^{\mu\nu\rho\sigma}}{\sqrt{\left|g\right|}}F_{\mu\nu}T_{\rho\sigma}^{b}+\dots,\label{eq:3+1_dimchiralEMT}
\end{eqnarray}
where $\dots$ means that only the most divergent terms are kept. Detailed calculations are relegated to Appendix~\ref{sec:chiral_energy_momentum_anomaly}.  
Finally, for the $\left(2+1\right)$-dimensional massive Dirac fermions, the
parity-odd effective action induced by torsion is
\begin{equation}
S_{\text{HLF}}=\frac{m^{2}\text{sgn}\left(m\right)}{8\pi}\int d^{3}x\epsilon^{\mu\nu\rho}e_{\mu}^{a}T_{\nu\rho}^{b}\eta_{ab},\label{eq:HLF_effective_action}
\end{equation}
where $m$ is the mass of Dirac fermions. Note that by direct perturbative
calculations \citep{hughes2011prl}, this effective action is divergent,
and it is proportional to the cutoff $\Lambda$. However, for reasons that
will be discussed in the next section, the HLF action here is written
in terms of the mass instead of the cutoff.

Clearly, the coefficients of the anomaly terms in Eq. (\ref{eq:chiral_em_anomaly_2}),
(\ref{eq:3+1_dimchiralEMT}), and Eq. (\ref{eq:HLF_effective_action})
are all dimensionful and divergent. One may be curious whether we can understand these
anomalies in a consistent way, or they are simply an artificial illness.
For Dirac fermions in curved space-time without torsion, it is known
that the chiral anomaly, the parity anomaly, and the gauge anomaly
consist of a dimensional ladder \citep{alvarez1985aop}. Namely, the
$\left(2n+1\right)$-dimensional parity anomaly can be obtained from the $\left(2n+2\right)$-dimensional
chiral anomaly by using the Atiyah-Patodi-Singer index theorem. Due
to the Callan-Harvey mechanism, the current inflow captured by the
$\left(2n+1\right)$-dimensional parity odd effective action (parity anomaly)
leads to the $\left(2n\right)$-dimensional gauge anomaly at the edges. By
employing the same idea, the torsional anomalies can be derived as
well \citep{parrikar2014prd}. However, it is also tempting to derive
these torsional anomalies from the single-body Hamiltonian, which
emphasizes the physical mechanism behind these torsional anomalies. Finally, it is worth mentioning that in Ref. \cite{Nissinen2019arxivNY}, the cut-off in the Nieh-Yan anomaly is argued to be proportional to the Fermi momentum in the $^{3}$He-$A$ superfluid. 

\section{Anomalies of the chiral energy-momentum tensor\label{sec:ChEM_anomaly}}

In this section, we shall study the anomalies encoded in the chiral energy-momentum tensor based on the Hamiltonian approach (boxes in the first row in Fig. \ref{fig:Dimensional_ladder}). First of all, we shall derive the $\left(1+1\right)$-dimensional chiral
energy-momentum anomaly from the single-body Hamiltonian. Then, both
the $\left(2+1\right)$-dimensional HLF effective action and the $\left(3+1\right)$-dimensional 
chiral energy-momentum anomaly can easily be obtained from the $\left(1+1\right)$-dimensional
chiral energy-momentum anomaly. Finally, we further show how to derive
the $\left(3+1\right)$-dimensional chiral energy-momentum anomaly from the
$\left(2+1\right)$-dimensional HLF effective action, which completes the
first row of the dimensional ladders for the torsional anomalies
shown in Fig. \ref{fig:Dimensional_ladder}.

\subsection{$\left(1+1\right)$-dimensional chiral energy-momentum anomaly from the single-body
Hamiltonian}

The Hamiltonian for the $\left(1+1\right)$-dimensional Dirac fermions is
\[
H=\left(-i\partial_{z}\right)\sigma^{3},
\]
where $\sigma^{3}$ is the Pauli matrix, i.e., $\sigma^{3}=\text{diag}\left(1,\thinspace-1\right)$
and the chirality of the fermions is labeled by $s=\pm1$. Then, we turn
on the vielbein. For simplicity, we assume that $e_{\mu}^{*a}=\delta_{\mu}^{a}+\delta_{3}^{a}\delta_{\mu}^{z}\Phi\left(t\right)$
with $\partial_{t}\Phi\ll1$ and $\Phi\ll1$. The corresponding torsional
electric field is given as ${T^{3}}_{E}=\partial_{t}e_{z}^{*3}-\partial_{z}e_{t}^{*3}=\partial_{t}\Phi$.
Because $e_{a}^{\mu}$ is defined from $e_{\mu}^{*a}$ by requiring
$e_{\mu}^{*a}e_{b}^{\mu}=\delta_{b}^{a}$, $e_{a}^{\mu}\simeq\delta_{a}^{\mu}-\delta_{z}^{\mu}\delta_{a}^{3}\Phi$.
The coupling between the Pauli matrices and the vielbein is $\sigma^{a}\rightarrow\sigma^{a}e_{a}^{\mu}$,
so the Hamiltonian becomes
\begin{equation}
H=\left[1-\Phi\left(t\right)\right]\left(-i\partial_{z}\right)\sigma^{3}.
\end{equation}
 Notice that $\partial_{t}\Phi\left(t\right)\ll1$, so one can obtain the dispersion relation, i.e.,
$\mathcal{E}_{s}=s\left(1-\Phi\right)p_{z}$ . In addition, the effective
Fermi velocity is now $v_{F}=\left(1-\Phi\right)$, so $v_{F}$ can
be tuned by changing the vielbein adiabatically. Hence, if we impose
a cutoff on the energy, i.e., $-\Lambda<\mathcal{E}_{s}<\Lambda$,
we can squeeze particles and thus momentum out of the system by tuning
$\Phi$, which leads to the energy-momentum anomaly for $s$-fermions.
To be more concrete, let us calculate the momentum density for $s$-fermions,
i.e.,
\begin{eqnarray}
\tau_{\left(c\right)3}^{st} & = & \int\frac{dp_{z}}{2\pi}n_{F}\left[\mathcal{E}_{s}\left(p_{z}\right)\right]p_{z}\nonumber \\
 & = & \frac{1}{\left(1-\Phi\right)^{2}}\int_{-\Lambda s}^{+\Lambda s}\frac{d\mathcal{E}_{s}}{2\pi}\frac{1}{\exp\left(\beta\mathcal{E}_{s}\right)+1}\mathcal{E}_{s}\nonumber \\
 & \overset{T=0}{=} & -\frac{s}{4\pi\left(1-\Phi\right)^{2}}\Lambda^{2},\label{eq:momentum_density}
\end{eqnarray}
where $s=\pm1$ stands for the chirality and $\mathcal{E}_{s}=s\left(1-\Phi\right)p_{z}$
is the dispersion relation. $\Lambda$ is the energy cutoff, i.e.,
$-\Lambda<\left(1-\Phi\right)p_{z}<\Lambda$, and it measures the
depth of the vacuum. $n_{F}\left(\mathcal{E}_{s}\right)=\left[\exp\left(\beta\mathcal{E}_{s}\right)+1\right]^{-1}$
is the Fermi-Dirac distribution, with $\beta=\frac{1}{T}$ being the inverse
of the temperature $T$ . Equation (\ref{eq:momentum_density})
tells us that the averaged momentum density of the occupied electrons
receives corrections from the vielbein. The energy cutoff is fixed
at $\Lambda$, so the changes in $\Phi$ will change the Fermi velocity
and thus the averaged momentum \cite{hughes2013prd}. As a consequence, there is an anomaly
in the momentum current for $s$-fermions.

From Eq. (\ref{eq:momentum_density}), the time derivative of the
chiral momentum density at the small $\Phi$ approximation is
\[
\partial_{t}\tau_{\left(c\right)3}^{5t}=-\frac{\Lambda^{2}}{\pi}\partial_{t}\Phi,
\]
or in a covariant form,
\begin{equation}
\nabla_{\mu}\tau_{\left(c\right)a}^{5\mu}=\frac{\Lambda^{2}}{2\pi}\eta_{ab}\frac{\epsilon^{\mu\nu}}{\sqrt{\left|g\right|}}{T^{b}}_{\mu\nu}+\dots,
\end{equation}
where $\dots$ stands for other terms appearing at the classical level.
This is similar to the result in Eq. (\ref{eq:chiral_em_anomaly_2})
up to a multiplier of $2$, but we can always rescale the cutoff to
match them. The physical mechanism of this anomaly can be straightforwardly
appreciated as follow. Let us consider the right-handed fermions $\left(s=+1\right)$
for concreteness. Since the chemical potential is set to zero, all
the negative-energy states are occupied. The only energy scale in
the system is the cutoff, so by dimensional analysis, the averaged
momentum must be proportional to the cutoff square, i.e., $\tau_{\left(c\right)3}^{st}\propto\Lambda^{2}$.
In the presence of vielbeins, i.e., $e_{z}^{*3}=1+\Phi$, the cutoff
for the momentum becomes $\left(1+\Phi\right)\Lambda$. Hence, the change
in momentum density is $\Delta\tau_{\left(c\right)3}^{st}\propto\Lambda^{2}-\left(1+\Phi\right)^2\Lambda^{2}=-2\Phi\Lambda^{2}$,
which implies that $\frac{\Delta\tau_{\left(c\right)3}^{st}}{\Delta t}\propto-\Lambda^{2}\partial_{t}\Phi$,
and this leads to the chiral energy-momentum anomaly.

\subsection{Torsional anomalies from the $\left(1+1\right)$-dimensional chiral energy-momentum
anomaly}

After obtaining the $\left(1+1\right)$-dimensional chiral energy-momentum
anomaly, one can straightforwardly obtain other torsional anomalies,
including the $\left(2+1\right)$-dimensional HLF effective action, the $\left(3+1\right)$-dimensional
chiral energy-momentum anomaly, and the Nieh-Yan anomaly.

As for the HLF effective action, this can be done by employing the
Callan-Harvey mechanism \citep{callan1985npb,hughes2013prd,parrikar2014prd}.
Consider the $\left(2+1\right)$-dimensional Chern insulators with chiral
zero modes at the edges. Then, the $\left(1+1\right)$-dimensional chiral
energy-momentum anomaly derived in last section implies the existence
of the energy-momentum anomalies for the chiral edge modes. This energy-momentum
anomaly can only come from the current inflow in the bulk, so the
corresponding energy-momentum current in the bulk must be
\begin{equation}
\tau_{\left(c\right)a}^{\mu}=\frac{m^{2}\text{sgn}\left(m\right)}{4\pi}\frac{\epsilon^{\mu\nu\rho}}{\sqrt{\left|g\right|}}\eta_{ab}{T^{b}}_{\nu\rho},\label{eq:2+1_energy_inflow}
\end{equation}
where $m$ is the gap of the $\left(2+1\right)$-dimensional Chern insulators
and it acts as the cut-off for the edges states, so $\Lambda$ in
Eq. (\ref{eq:chiral_em_anomaly_2}) is replaced by $m$. In addition,
$\text{sgn}\left(m\right)$ is from the chirality of the edges modes.
Note that we are considering the continuum model in the bulk rather
than the tight-binding model, so the prefactor is $\text{sgn}\text{\ensuremath{\left(m\right)}}$.
Alternatively, if the tight-binding model is considered, then $\text{sgn}\left(m\right)$
should be replaced by $1-\text{sgn}\left(m\right)$. This difference comes from different regularization schemes.

From Eq. (\ref{eq:2+1_energy_inflow}), one can easily write down
the corresponding effective action, i.e.,
\begin{equation}
S_{\text{HLF}}=\frac{m^{2}\text{sgn}\left(m\right)}{8\pi}\int d^{3}x\epsilon^{\mu\nu\rho}e_{\mu}^{*a}{T^{b}}_{\nu\rho},
\end{equation}
where the canonical energy-momentum tensor is defined as $\tau_{\left(c\right)a}^{\mu}=\frac{-1}{\sqrt{\left|g\right|}}\frac{\delta S}{\delta e_{\mu}^{*a}}|_{\omega_{ab\mu}}$
with the spin connection $\omega_{ab\mu}$ kept fixed. Hence, we have
recovered the $\left(2+1\right)$-dimensional HLF effective action from the
$\left(1+1\right)$-dimensional chiral energy-momentum anomaly.

Interestingly, the $\left(3+1\right)$-dimensional Nieh-Yan anomaly can be
derived from the $\left(1+1\right)$-dimensional chiral energy-momentum anomaly
as well. Consider Weyl fermions under a specific configuration of
vielbeins, i.e., $e_{\mu}^{*a}=\delta_{\mu}^{a}+w_{\mu}^{a}$ and
$w_{\mu}^{a}=\frac{1}{2}\delta_{3}^{a}{\tilde{T}^{3}}_{B}\left(0,\thinspace-y,\thinspace x,\thinspace0\right),\text{\ensuremath{{\tilde{T}^{3}}_{B}>0}}$,
which means that the torsional magnetic fields are applied along the
$z$ direction. For simplicity, the spin connection is set to zero.
Then, the effective Hamiltonian for Weyl fermions with $s$-chirality
is

\begin{eqnarray}
H_{s} & = & s[p_{z}\sigma^{3}+\left(\hat{p}_{x}+\frac{1}{2}{\tilde{T}^{3}}_{B}yp_{z}\right)\sigma^{1}\nonumber \\
 &  & +\left(\hat{p}_{y}-\frac{1}{2}{\tilde{T}^{3}}_{B}xp_{z}\right)\sigma^{2}],
\end{eqnarray}
where $p_{z}$ is a good quantum number and $s=\pm1$ denotes the
chirality. This Hamiltonian looks like Weyl fermions under magnetic
fields ${\tilde{T}^{3}}_{B}$ with charge $p_{z}$. The dispersion
relation can be straightforwardly derived, i.e.,

\begin{equation}
\mathcal{E}_{s}=\begin{cases}
\begin{array}{c}
s\left|p_{z}\right|\\
\pm\sqrt{p_{z}^{2}+2\left|n{\tilde{T}^{3}}_{B}p_{z}\right|}
\end{array} & \begin{array}{c}
n=0\\
\left|n\right|\geq1
\end{array}\end{cases}.\label{eq:torsional_Landau_level}
\end{equation}
Compared to the magnetic case, the lowest torsional Landau levels
are $s\left|p_{z}\right|$ rather than $sp_{z}$, which is because the charge of the torsional magnetic fields is $p_{z}$ and it
contributes an extra minus sign when $p_{z}<0$. Notice that only
the lowest torsional Landau levels can distinguish Weyl fermions with
different chiralities. Thus, the $\left|n\right|>1$ torsional Landau
levels contribute equally to the $s$-Weyl fermion current. That is,
as far as the chiral anomaly is concerned, only the lowest torsional
Landau levels matter. Consequently, the system is effectively reduced
from $\left(3+1\right)$-dimensional to $\left(1+1\right)$-dimensional and the corresponding
effective $\left(1+1\right)$-dimensional Lagrangian is
\begin{equation}
\mathcal{L}=\frac{1}{2}\left(\bar{\psi^{\prime}}\Gamma^{\mu}i\partial_{\mu}\psi^{\prime}-\bar{\psi^{\prime}}i\overleftarrow{\partial}_{\mu}\Gamma^{\mu}\psi^{\prime}\right),
\end{equation}
where $\Gamma^{\mu}$ is the $\left(1+1\right)$-dimensional gamma matrix,
$\psi^{\prime}\left(p_{z}\right)=\left(\begin{array}{c}
\psi_{R}\\
\psi_{L}
\end{array}\right)$ for $p_{z}>0$, and $\psi^{\prime}\left(p_{z}\right)=\left(\begin{array}{c}
\psi_{L}\\
\psi_{R}
\end{array}\right)$ for $p_{z}<0$. The $\left(3+1\right)$-dimensional chiral current can be
defined in terms of the $\left(1+1\right)$-dimensional Dirac fermions, i.e.,
\begin{equation}
j^{5\mu}=\left(\bar{\psi^{\prime}}\Gamma^{\mu}\Gamma^{5}p_{z}\psi^{\prime}\right)\left(\frac{{\tilde{T}^{3}}_{B}}{2\pi}\right),
\end{equation}
where ${\tilde{T}^{3}}_{B}$ is the torsional magnetic field, $\frac{1}{2\pi}\left(\left|p_{z}\right|{\tilde{T}^{3}}_{B}\right)$
is from the torsional Landau level degeneracy, and the chirality of
$\psi^{\prime}$ is twisted by $\text{sgn}\left(p_{z}\right)$. Especially,
$\left(\bar{\psi^{\prime}}\Gamma^{\mu}\Gamma^{5}p_{z}\psi^{\prime}\right)$
is the $\left(1+1\right)$-dimensional chiral energy-momentum density $\tau_{\left(c\right)z}^{5\left(1+1\right)\mu}$.
Hence,  $\nabla_{\mu}j^{5\mu}=\left(\nabla_{\mu}\tau_{\left(c\right)z}^{5\left(1+1\right)\mu}\right){\tilde{T}^{3}}_{B},$
where $\nabla_{\mu}\tau_{\left(c\right)a}^{5\left(1+1\right)\mu}$
is the $\left(1+1\right)$-dimensional chiral energy-momentum anomaly. Consequently,
the Nieh-Yan anomaly in Eq. (\ref{eq:Nieh_Yan}) can be derived from
the $\left(1+1\right)$-dimensional chiral energy-momentum anomaly.

As for the $\left(3+1\right)$-dimensional chiral energy-momentum anomaly in Eq. \ref{eq:3+1_dimchiralEMT}, it involves a term mixing the torsion tensor and the electromagnetic tensor. This anomaly
can also be obtained from the $\left(1+1\right)$-dimensional chiral energy-momentum
anomaly. Namely, under external magnetic fields, the $\left(3+1\right)$-dimensional
Dirac fermions are effectively reduced to the $\left(1+1\right)$-dimensional
ones consisting of the lowest Landau levels. The $\left(3+1\right)$-dimensional
chiral energy-momentum anomaly can be written as $\nabla_{\mu}\tau_{\left(c\right)a}^{5\left(3+1\right)\mu}\propto\nabla_{\mu}\tau_{\left(c\right)a}^{5\left(1+1\right)\mu}B$,
which coincides with the anomaly in Eq. (\ref{eq:3+1_dimchiralEMT}).
So far, we have shown that most known torsional anomalies can be
understood from the $\left(1+1\right)$-dimensional chiral energy-momentum
anomaly. It is natural to ask the relation between the $\left(2+1\right)$-dimensional
HLF effective action and the $\left(3+1\right)$-dimensional torsional anomalies.

\subsection{$\left(3+1\right)$-dimensional chiral energy-momentum anomaly from the $\left(2+1\right)$-dimensional
HLF effective action }

In this section, we shall reveal the relation between the HLF effective
action and the $\left(3+1\right)$-dimensional chiral energy-momentum anomaly
based on the single-body Hamiltonian.

The Hamiltonian for the $\left(3+1\right)$-dimensional right-handed Weyl
fermions under torsion is
\begin{equation}
H=H_{\perp}\left(\hat{p}_{x},\thinspace\hat{p}_{y},\thinspace e_{\perp\mu}^{*a}\right)+p_{z}\sigma^{3},
\end{equation}
where the background curvature is assumed to be zero, so we can set
the spin connection to zero. $H_{\perp}\left(\hat{p}_{x},\thinspace\hat{p}_{y},\thinspace e_{\perp\mu}^{*a}\right)$
is the Hamiltonian for the $\left(2+1\right)$-dimensional massless Dirac
fermions with vielbeins and $e_{\perp\mu}^{*a}=e_{\perp\mu}^{*a}\left(x,\thinspace y\right)$
is the vielbein in the space perpendicular to $z$. In addition, $p_{z}$
is a good quantum number. Note that $p_{z}\sigma^{3}$ looks like
a mass term for the $\left(2+1\right)$-dimensional Dirac fermions. So from
the HLF effective action, the momentum density at a given momentum $p_{z}$
is
\begin{equation}
\rho_{\left(c\right)a}^{\left(+\right)t}\left(p_{z}\right)=\frac{p_{z}^{2}\text{sgn}\left(p_{z}\right)}{4\pi}\eta_{ab}{T^{b}}_{B},
\end{equation}
where the superscript $\left(+\right)$ denotes the chirality, the
mass term $m^{2}\text{sgn}\left(m\right)$ is replaced by $p_{z}^{2}\text{sgn}\left(p_{z}\right)$
and ${T^{a}}_{B}=-\left(\partial_{x}e_{y}^{*a}-\partial_{y}e_{x}^{*a}\right)$
is the torsional magnetic fields. Now let us further turn on the electric
fields, i.e., $p_{z}\rightarrow p_{z}-A_{z}\left(t\right)$, where
$\partial_{t}A_{z}\ll1$.
Thus, the momentum density at given momentum $p_{z}$ becomes
\begin{equation}
\rho_{\left(c\right)a}^{\left(+\right)t}\left(p_{z}-A_{z}\right)=\frac{\left(p_{z}-A_{z}\right)^{2}\text{sgn}\left(p_{z}-A_{z}\right)}{4\pi}\eta_{ab}{T^{b}}_{B},
\end{equation}
which implies that the momentum density of the right-handed Weyl fermions
can be tuned by varying $A_{z}$ adiabatically. Hence, the momentum
for the right-handed Weyl fermions can be pumped up from the Dirac
sea by changing $A_{z}$ . Notice that the effective masses for the
right-handed and left-handed Weyl fermions always have opposite signs.
So the increasing of the momentum density of the right-handed Weyl
fermions is accompanied by the decreasing of the momentum density
of left-handed ones. Consequently, there is an anomaly in the $\left(3+1\right)$-dimensional
chiral energy-momentum tensor, but not the energy-momentum tensor.

Now let us calculate the total momentum density by integrating over
$p_{z}$. The total momentum density for right-handed Weyl fermions
is given as

\begin{eqnarray}
\tau_{\left(c\right)a}^{\left(+\right)t} & = & \int_{-\Lambda}^{+\Lambda}\frac{dp_{z}}{2\pi}\rho_{\left(c\right)a}^{t}\nonumber \\
 & = & -\frac{\Lambda^{2}}{4\pi^{2}}\eta_{ab}A_{z}{T^{b}}_{B}.
\end{eqnarray}
 This implies that
\begin{equation}
\partial_{t}\tau_{\left(c\right)a}^{5t}=-\frac{\Lambda^{2}}{2\pi^{2}}\eta_{ab}E_{z}{T^{a}}_{B}+\dots,
\end{equation}
and its covariant form is the $\left(3+1\right)$-dimensional chiral energy-momentum
anomaly given in Eq. (\ref{eq:3+1_dimchiralEMT}). Hence, we have
completed the first part of the dimensional ladder shown in Fig. \ref{fig:Dimensional_ladder}.

\section{Anomalies of the chiral current \label{sec:Ch_current_anomaly}}

In the last section, we completed the dimensional ladder for the
chiral energy-momentum anomaly. In this section, we shall show that
there is an analog ladder for the anomalies encoded in the chiral
current as well, which is shown in the second row in Fig. \ref{fig:Dimensional_ladder}.
Namely, the Nieh-Yan anomaly can be understood from the $\left(2+1\right)$-dimensional
parity anomaly, and the $\left(2+1\right)$-dimensional parity anomaly can
be obtained from the $\left(1+1\right)$-dimensional chiral anomaly.

The relation between the $\left(2+1\right)$-dimensional parity anomaly and
the $\left(1+1\right)$-dimensional chiral anomaly is well known in the context
of the integer quantum Hall effect, which is used to understand the bulk-edge
correspondence. For completeness, we shall briefly review it here.
The covariant form of the Hall current in $\left(2+1\right)$ dimensions is 
\begin{equation}
j^{\mu}=\frac{\text{sgn}\left(m\right)}{8\pi}\epsilon^{\mu\rho\sigma}F_{\rho\sigma},\label{eq:hall_current}
\end{equation}
where $m$ is the mass of the $\left(2+1\right)$-dimensional Dirac fermions
and $\text{sgn}\left(m\right)$ is from the chirality of the edge
modes. In addition, the spatial components of the current above can
be recast in a more inspiring form, i.e., $\boldsymbol{j}=\frac{\text{sgn}\left(m\right)}{4\pi}\hat{z}\times\boldsymbol{E}$.
If we consider the tight-binding model in the bulk instead, then the
Hall current becomes $\boldsymbol{j}=\frac{1}{2\pi}\left[\frac{1-\text{sgn}\left(m\right)}{2}\right]\hat{z}\times\boldsymbol{E}$, 
and $\frac{1}{2}\left[1-\text{sgn}\left(m\right)\right]$ is the
Chern number in the bulk. So for the topologically non-trivial phase,
$\frac{1-\text{sgn}\left(m\right)}{2}=1$, we have obtained the quantized
Hall conductance, i.e., $\sigma_{H}=\frac{1}{2\pi}$. From the Hall
current in Eq. (\ref{eq:hall_current}), one can write down the parity
odd effective action (the parity anomaly), i.e.,
\begin{equation}
S=\frac{\text{sgn}\left(m\right)}{16\pi}\int d^{3}x\epsilon^{\mu\rho\sigma}A_{\mu}F_{\rho\sigma}.
\end{equation}
In addition, Eq. (\ref{eq:hall_current}) also tells us that there
is current flowing toward the edges, which leads to the current nonconservation
at the edges or the $\left(1+1\right)$-dim gauge anomaly.

Now let us turn to the Nieh-Yan anomaly and its relation to the parity
anomaly. The Hamiltonian for the right-handed Weyl fermions is
\begin{equation}
H=H_{\perp}\left(\hat{p}_{x},\,\hat{p}_{y};\,e_{\perp\mu}^{*a}\right)+p_{z}\sigma^{z},\label{eq:2+1weyl_NiehYan}
\end{equation}
where $p_{z}$ is a good quantum number and $H_{\perp}$ is the Hamiltonian
for the $\left(2+1\right)$-dimensional Dirac fermions under external vielbeins,
i.e., $e_{\perp a}^{\mu}=\delta_{a}^{\mu}-w_{\perp a}^{\mu}\left(x,\,y\right)$.
For concreteness, we further assume that $H_{\perp}$ is given as
\begin{equation}
H_{\perp}\left(\hat{p}_{x},\,\hat{p}_{y};\,e_{\perp\mu}^{*a}\right)=\left(\hat{p}_{x}-w_{1}^{z}p_{z}\right)\sigma^{x}+\left(\hat{p}_{y}-w_{2}^{z}p_{z}\right)\sigma^{z}.
\end{equation}
Hence, Eq. (\ref{eq:2+1weyl_NiehYan}) looks like $\left(2+1\right)$-dimensional
massive Dirac fermions (the mass term is $p_{z}\sigma^{3}$) under
external electromagnetic field $w_{i}^{z}$ with charge $p_{z}$.
From the parity anomaly, one can straightforwardly read off the Hall
current density  at given momentum $p_{z}$, i.e.,
\begin{equation}
\rho^{\left(+\right)}\text{\ensuremath{\left(p_{z}\right)}}=\frac{1}{4\pi}p_{z}\text{sgn}\left(p_{z}\right){T^{3}}_{B},
\end{equation}
where ${T^{3}}_{B}$ is the torsional magnetic field, i.e., ${T^{3}}_{B}=\partial_{x}w_{y}^{3}-\partial_{y}w_{x}^{3}$.
Now we further turn on the torsional electric fields, i.e., ${T^{3}}_{E}=\partial_{t}e_{z}^{*3}-\partial_{z}e_{t}^{*3}$.
For simplicity, we set $e_{t}^{*3}=0$ and $e_{z}^{*3}=1+\Phi\left(t\right)$
with $\Phi$ being a slowly varying field. Hence, the Hall density becomes
\begin{equation}
\rho^{\left(+\right)}\left(p_{z}\right)=\frac{1}{4\pi}p_{z}\text{sgn}\left[\left(1-\Phi\right)p_{z}\right]{T^{3}}_{B},
\end{equation}
where $e_{z}^{*3}$ modifies only the mass term, and not the charge
of the effective electromagnetic fields $w_{i}^{z}$. The total current
density can be derived by summing over the momentum,
\begin{eqnarray}
j^{\left(+\right)0} & = & \int\frac{dp_{z}}{2\pi}\rho^{\left(+\right)}\left(p_{z}\right)\nonumber \\
 & = & \frac{1}{8\pi^{2}\left(1-\Phi\right)^{2}}\int_{-\Lambda}^{\Lambda}d\mathcal{E}\mathcal{E}\text{sgn}\mathcal{E}{T^{3}}_{B}\nonumber \\
 & \simeq & \frac{1}{4\pi^{2}}\Lambda^{2}\Phi{T^{3}}_{B},
\end{eqnarray}
where $\mathcal{E}\equiv\left(1-\Phi\right)p_{z}$ is the energy and
we have imposed a cutoff $\Lambda$ upon the energy. By recasting the
result above to a covariant form, one can obtain the Nieh-Yan anomaly
in Eq. (\ref{eq:Nieh_Yan}). This completes the second part of the
dimensional ladder shown in Fig. \ref{fig:Dimensional_ladder}.

It is interesting to notice that the $\left(3+1\right)$-dimensional chiral
energy-momentum anomaly involves the mixing term between the torsion
and the electromagnetic tensor, while the Nieh-Yan anomaly involves only the
torsion tensor. Furthermore, if we regard the $\left(3+1\right)$-dimensional
Weyl fermions as the boundary chiral modes of some $\left(4+1\right)$-dimensional
topological phases of matter, then from the Callan-Harvey mechanism,
one can obtain the corresponding $\left(4+1\right)$-dimensional effective
action \citep{parrikar2014prd}
\begin{equation}
S_{\left(4+1\right)}=\frac{m^{2}\text{sgn}\left(m\right)}{64\pi^{2}}\int d^{5}x\epsilon^{\mu\nu\rho\sigma\delta}A_{\mu}{T^{a}}_{\nu\rho}{T^{b}}_{\sigma\delta}\eta_{ab},
\end{equation}
whose response charge current and energy-momentum current are the
bulk origin of the boundary Nieh-Yan anomaly and chiral energy-momentum
anomaly, respectively.

\section{Summary and discussion \label{sec:summary_dis}}

The dimensional ladder for torsional anomalies is constructed based
on the single-body Hamiltonian. We have shown how to obtain the $\left(2+1\right)$-dimensional
HLF effective action, the $\left(3+1\right)$-dimensional Nieh-Yan anomaly
and chiral energy-momentum anomaly from the $\left(1+1\right)$-dimensional
chiral energy-momentum anomaly. In addition, we have also clarified
the relation between the $\left(2+1\right)$-dimensional HLF effective action
and the $\left(3+1\right)$-dimensional chiral energy-momentum anomaly, the
$\left(2+1\right)$-dimensional parity anomaly and the $\left(3+1\right)$-dimensional
Nieh-Yan anomaly. Our work has provided a complete physical picture
for various torsional anomalies in various dimensions.

Recently, it was pointed out in Ref. \citep{nissinen2019arxivSF,nissinen2019arxivWM,huang2019arxiv}
that, at finite temperature, there is an extra thermal term in the
Nieh-Yan anomaly as well, where the cut-off is replaced by the temperature.
Hence, based on the dimensional ladder obtained here, it is natural to
conjecture that all the torsional anomalies in Fig. \ref{fig:Dimensional_ladder} that are cut-off dependence
receive thermal corrections as well. That is, the dimensionful
parameter is replaced by the temperature. Specifically, for the $\left(1+1\right)$-dimensional chiral energy-momentum anomaly calculated in Eq. (\ref{eq:momentum_density}), if we use the finite-temperature Fermi-Dirac distribution function instead of the zero-temperature one, then there is an extra thermal term proportional to the temperature square.  If this conjecture is true,
then the $\left(2+1\right)$-dimensional thermal Hall effect can be straightforwardly
understood from the $\left(2+1\right)$-dimensional thermal HLF effective
action in the bulk, or $\left(1+1\right)$-dimensional energy-momentum
anomaly at the edges. In addition, the recently observed negative magneto-thermoelectric resistance reported in Ref. \citep{gooth2017nature} can be understood based on this conjecture as well. Namely, if there is an extra thermal term in the  $(3+1)$-dimensional chiral energy-momentum anomaly equation  in Eq. (\ref{eq:3+1_dimchiralEMT}), i.e., $\nabla_\mu \tau^{5\mu}_{a}\propto T^2 \frac{\epsilon^{\mu\nu\rho\sigma}}{\sqrt{|g|}}\eta_{ab}F_{\mu\nu}T^{b}_{\rho\sigma}+\dots$, then by considering the inter valley scatterings, in the steady states, the chiral chemical potential is proportional to $\epsilon^{\mu\nu\rho\sigma}\eta_{ab}F_{\mu\nu}T^{a}_{\rho\sigma}$, or  $T(\boldsymbol{B}\cdot \boldsymbol{\nabla}{T})$ for $a=0$. Together with the chiral magnetic effect, one can thus obtain the negative magnetothermoelectric resistance. However, here, we shall restrict ourselves to
the zero-temperature limit and leave the conjecture for future study.

\section{Acknowledgments}

The authors wish to thank B. Bradlyn for insightful
discussions. Z.-M.-H. and M.-S. were not directly supported by any funding agencies,
but this work would not be possible without resources provided by
the Department of Physics at the University of Illinois at Urbana-Champaign. B. H. was supported by  ERC Starting Grant No. 678795 TopInSy.

\appendix

\section{Fujikawa's approach to the Nieh-Yan anomaly \label{sec:Nie_Yan_fujikawa}}

In this appendix, we shall provide a detailed derivation of the torsion
induced chiral anomaly. For convenience, we shall focus on the Euclidean
space-time. The convention of the Wick rotation is

\begin{equation}
t=-i\tau,\thinspace\gamma_{E}^{0}=\gamma^{0},\thinspace \thinspace\thinspace\gamma_{E}^{i}=-i\gamma^{i},
\end{equation}
where $\tau$ is the Euclidean time. The subscript $E$ stands for
the Euclidean spacetime, which will be neglected without causing
any confusion, because in this appendix we will focus on the Euclidean
space-time.

The Jacobian $J\left[\theta\right]$ of the chiral transformation
$\psi\rightarrow e^{i\theta\gamma_5}\psi$ is known to be \citep{fujikawa2004oxford}

\begin{equation}
\ln J\left[\theta\right]=-2i\theta \text{Tr}\gamma_{5},
\end{equation}
where $\text{Tr}$ is the trace of both the internal indices and the
spacetime coordinates. $\text{Tr}\gamma_{5}$ is obviously divergent,
so we introduce the following regulator,

\[
\exp\left(\frac{\slashed{\mathcal{D}}\slashed{\mathcal{D}}}{\Lambda^{2}}\right),
\]
where
\[
\slashed{\mathcal{D}}=e_{a}^{\mu}\gamma^{a}\left(\partial_{\mu}+\frac{1}{2}\omega_{ab\mu}\sigma^{ab}+\frac{1}{2}{T^{\rho}}_{\mu\rho}\right)
\]
and

\[
\overleftarrow{\slashed{\mathcal{D}}}=\left(\partial_{\mu}-\frac{1}{2}\omega_{ab\mu}\sigma^{ab}+\frac{1}{2}{T^{\rho}}_{\mu\rho}\right)e_{a}^{\mu}\gamma^{a}
\]
are skew Hermitian. Note that $\slashed{D}$ is neither Hermitian
nor skew Hermitian, so we use $\mathcal{D}$ rather than $D$ and
the extra term $\frac{1}{2}{T^{\rho}}_{\mu\rho}$ in $\slashed{\mathcal{D}}$
and $\overleftarrow{\slashed{\mathcal{D}}}$ is canceled in the action.

Then, for $\slashed{\mathcal{D}}\slashed{\mathcal{D}}$, there is

\begin{eqnarray*}
 &  & \left(\slashed{\mathcal{D}}\right)^{2}\\
 & = & \delta^{ab}\mathcal{D}_{a}\mathcal{D}_{b}+\sigma^{ab}\left(-{T^{d}}_{ab}\mathcal{D}_{d}+\frac{1}{2}\Omega_{cd,\thinspace ab}\sigma^{cd}\right)\\
 &  & +\sigma^{ab}\left(\partial_{a}{T^{\rho}}_{b\rho}-\mathcal{\partial}_{b}{T^{\rho}}_{a\rho}\right)
\end{eqnarray*}
where we have used

\[
\left[D_{a},\thinspace D_{b}\right]\psi=\left(-e_{a}^{\mu}e_{b}^{\nu}{T^{d}}_{\mu\nu}D_{d}+\frac{1}{2}e_{a}^{\mu}e_{b}^{\nu}\Omega_{\mu\nu}\right)\psi,
\]
and thus

\begin{eqnarray*}
 &  & \left[\mathcal{D}_{a},\thinspace\mathcal{D}_{b}\right]\psi\\
 & = & \left[-{T^{d}}_{ab}D_{d}+\frac{1}{2}\Omega_{cd,\thinspace ab}\sigma^{cd}\right]\psi.\\
 &  & +\left(\partial_{a}{T^{\rho}}_{b\rho}-\partial_{b}{T^{\rho}}_{a\rho}\right)\psi.
\end{eqnarray*}

Because we are most interested in the torsion, for simplicity,
let us set the curvature $\Omega_{ab}=\frac{1}{2}\Omega_{ab\mu\nu}dx^{\mu}\wedge dx^{\nu}$
to zero. Hence, we can choose such a gauge that the spin connection
is zero. In addition, in the presence of electromagnetic fields, 

\begin{eqnarray*}
\left(\slashed{\mathcal{D}}\right)^{2} & \simeq & \delta^{ab}\mathcal{D}_{a}\mathcal{D}_{b}+\sigma^{ab}\left(-{T^{d}}_{ab}D_{d}+iF_{ab}\right),
\end{eqnarray*}
where $\simeq$ is because that we are most interested in the most
divergent terms, $\left(\partial_{a}T^{\rho}_{b\rho}-\partial_{b}T^{\rho}_{a\rho}\right)$
is neglected. Notice that in the absence of curvature, 
\[
\left\{ \mathcal{D}_{a},\thinspace\mathcal{D}_{b}\right\}=\partial_{b}\left(iA_{a}+\frac{1}{2}T^{\rho}_{a\rho}\right)+(a\leftrightarrow b)+\dots,
\]
so we can choose such a gauge fixing condition that $\partial_{a}A_{b}+\partial_{b}A_{a}=0.$
In addition, we shall neglect the terms from
${T^{\rho}}_{\mu\rho}$ hereafter, which is reasonable because we are interested in only the most divergent terms.

Then, 
\begin{widetext}
\begin{eqnarray*}
 &  & -2i\text{Tr}\left(\gamma^{5}e^{\Lambda^{-2}\slashed{\mathcal{D}}\slashed{\mathcal{D}}}\right)\\
 & = & -2i\sum_{n}\frac{1}{n!}\int d^{d}x\int\frac{d^{d}k}{\left(2\pi\right)^{d}}e^{-\Lambda^{-2}k^{2}}\text{tr}\left(\gamma^{5}\sigma^{a_{1}b_{1}}\sigma^{a_{2}b_{2}}\dots\sigma^{a_{n}b_{n}}\right)\prod_{i=1}^{n}\left(-i\Lambda^{-2}{T^{d}}_{a_{i}b_{i}}k_{d}+i\Lambda^{-2}F_{a_{i}b_{i}}\right)^{n}\\
 & \rightarrow & -2i\sum_{n}\frac{1}{n!}\int d^{d}x\int\frac{d^{d}k}{\left(2\pi\right)^{d}}e^{-\Lambda^{-2}k^{2}}\left(-2\Lambda^{-2}{T^{d}}k_{d}+2\Lambda^{-2}F\right)^{n}\\
 & = & -2i\sum_{n}\sum_{p=0,\thinspace2,\thinspace4\dots}C_{n}^{p}\frac{1}{n!}\int d^{d}x\int\frac{d^{d}k}{\left(2\pi\right)^{d}}e^{-\left(\Lambda^{2}\right)^{-1}k^{2}}\left(-2\left(\Lambda^{2}\right)^{-1}{T^{d}}k_{d}\right)^{p}\left(2\left(\Lambda^{2}\right)^{-1}F\right)^{n-p}\\
 & = & -2i\sum_{p\in\text{even}}\int d^{d}x\left[\frac{\left(\Lambda^{2}\right)^{\frac{1}{2}\left(d+p\right)}}{2^{\frac{1}{2}p}\left(4\pi\right)^{d/2}}\sum_{\left\{ \sigma\right\} }\left(\delta_{\sigma\left(b_{1}\right)\sigma\left(b_{2}\right)}\delta_{\sigma\left(b_{3}\right)\sigma\left(b_{4}\right)}\dots\right)\right]\frac{\left(-2\Lambda^{-2}{T^{b}}\right)^{p}}{p!}e^{2\left(\Lambda^{2}\right)^{-1}F}.
\end{eqnarray*}
The $\rightarrow$ symbol in the third line is because we have replaced
$-i\sigma^{a_{1}b_{1}}\left({T^{d}}_{a_{1}b_{1}}+F_{a_{1}b_{1}}\right)$
with $2{T^{d}}+2F$ and ${T^{d}}=\frac{1}{2}{T^{d}}_{\mu\nu}dx^{\mu}\wedge dx^{\nu}$,
$F=\frac{1}{2}F_{\mu\nu}dx^{\mu}\wedge dx^{\nu}$, which is because
of the following identity
\[
\text{tr}\left(\gamma^{5}\sigma^{a_{1}b_{1}}\sigma^{a_{2}b_{2}}\dots\sigma^{a_{n}b_{n}}\right)=(-1)\left(-i\right)^{n}\epsilon^{a_{1}b_{1}\dots a_{n}b_{n}}.
\]
In the fourth line, we perform the expansion $\left(a+b\right)^{n}=\sum_{m}C_{n}^{m}a^{m}b^{\left(n-m\right)}.$
In the last line, we use the following identity
\begin{eqnarray*}
 &  & \int\frac{d^{d}k}{\left(2\pi\right)^{d}}\exp\left(-k_{a}\Lambda^{-2}\delta_{ab}k_{b}\right)k_{a_{1}}k_{a_{2}}\dots k_{a_{2m}}\\
 & = & \left(\frac{1}{2}\right)^{m}\frac{\left(\Lambda^{2}\right)^{d/2+m}}{\left(4\pi\right)^{d/2}}\left(\sum_{\left\{ \sigma\right\} }\delta_{\sigma\left(a_{1}\right)\sigma\left(a_{2}\right)}\dots\right),
\end{eqnarray*}
where $\sum_{\left\{ \sigma\right\} }$ means a sum over all possible
permutations.
\end{widetext}

Notice that $T^{b_{1}}$ and $T^{b_{2}}$ in $\left(-2\Lambda^{-2}{T^{b}}\right)^{p}$
commute, so the anomaly above can be recast as

\begin{eqnarray}
 &  & -2i(-1)\int\sum_{p\in\text{even}}\left[\frac{\left(\Lambda^{2}\right)^{\frac{1}{2}\left(d+p\right)}\left(p-1\right)!!}{2^{\frac{1}{2}p}\left(4\pi\right)^{d/2}}\right]\frac{\left(4\Lambda^{-4}{T^{c_{1}}}{T^{c_{2}}}\delta_{c_{1}c_{2}}\right)^{p/2}}{p!}e^{\frac{2F}{\Lambda^{2}}}\nonumber \\
 &  &= -2i(-1)\int d^{d}x\frac{\left(\Lambda^{2}\right)^{\frac{1}{2}d}}{\left(4\pi\right)^{d/2}}e^{\Lambda^{-2}{T^{c_{1}}}{T^{c_{2}}}\delta_{c_{1}c_{2}}}e^{2\Lambda^{-2}F},\nonumber \label{eq:anomaly}
\end{eqnarray}
where $\sum_{\sigma}\left(\delta_{\sigma\left(b_{1}\right)\sigma\left(b_{2}\right)}\delta_{\sigma\left(b_{3}\right)\sigma\left(b_{4}\right)}\dots\right)$
contains $\left(p-1\right)!!$ terms and $\left(-1\right)^{p}=1$
because that $p$ is an even integer. That is, we have obtained
\begin{equation}
-2i\text{Tr}\gamma_{5}=-2i(-1)\int\frac{\left(\Lambda^{2}\right)^{\frac{1}{2}d}}{\left(4\pi\right)^{d/2}}e^{\Lambda^{-2}{T^{c_{1}}}{T^{c_{2}}}\delta_{c_{1}c_{2}}}e^{2\Lambda^{-2}F}+\dots,
\end{equation}
where $e^{2\Lambda^{-2}F}$ is the Chern character and ${T^{c_{1}}}{T^{c_{2}}}\eta_{c_{1}c_{2}}$
reminds us of the Nieh-Yan term, i.e., $T^{a}\wedge T_{a}-e^{a}\wedge e^{b}\wedge\Omega_{ab}$.
Thus, it is natural to conjecture that the full anomaly polynomial should be
\[
-2i\text{Tr}\gamma_{5}=2i\int\frac{\left(\Lambda^{2}\right)^{\frac{1}{2}d}}{\left(4\pi\right)^{d/2}}e^{2\Lambda^{-2}\left(T^{a}\wedge T^{b}\delta_{ab}-e^{a}\wedge e^{b}\wedge\Omega_{ab}\right)}e^{2\Lambda^{-2}F}\hat{A}\left(\Omega\right),
\]
where $\hat{A}\left(\Omega\right)$ is the Dirac genus.

\section{Chiral energy-momentum anomaly \label{sec:chiral_energy_momentum_anomaly}}

In this appendix, we shall provide the derivation of the chiral energy-momentum
anomaly in detail. We first derive the constraints from the covariant
diffeomorphism. After that, the anomaly in the zero-curvature limit
is also derived by using Fujikawa's method.

\subsection{Covariant diffeomorphism and the corresponding current}

Consider an action $S\left(\bar{\psi},\:\psi,\:e_{\mu}^{*a},\:\omega_{ab\mu}\right)$
with both local Lorentz symmetry and local Einstein symmetry. The
local Lorentz transformation is given as
\[
\delta e_{\mu}^{*a}=-{\alpha^{a}}_{b}e_{\mu}^{*b},
\]
\[
\delta\omega_{ab\mu}=D_{\mu}\alpha_{ab},
\]
and
\[
\delta\psi=-\frac{1}{2}\alpha_{ab}\sigma^{ab}\psi,
\]
where $\omega_{ab\mu}$ transforms as the gauge field for the Lorentz
group as we expected. The local Einstein transformation is
\[
\delta e_{\mu}^{*a}=\mathcal{L}_{\xi}e_{\mu}^{*a},
\]
\[
\delta\omega_{ab\mu}=\mathcal{L}_{\xi}\omega_{ab\mu},
\]
and
\[
\delta\psi=\xi^{\mu}\partial_{\mu}\psi,
\]
where $\mathcal{L}_{\xi}$ is the Lie derivative of the vector $\xi=\xi^{\mu}\partial_{\mu}$.
The local Einstein transformation is usually called diffeomorphism
as well. By performing a local Einstein transformation and then a
local Lorentz transformation with $\alpha_{ab}=\left(i_{\xi}\omega\right)_{ab}$,
one can obtain the covariant diffeomorphism, i.e.,

\begin{eqnarray}
\delta e_{\mu}^{*a} & = &- {T^{a}}_{\mu\nu}\xi^{\nu}+\nabla_{\mu}\xi^{a},\nonumber \\
\delta\omega_{ab\mu} & = & -\Omega_{ab\mu\nu}\xi^{\nu}\nonumber \\
\delta\psi & = & \xi^{\mu}D_{\mu}\psi
\end{eqnarray}

Then, the variation of the action

\begin{eqnarray*}
\delta S & = & \int d^{d}x\sqrt{\left|g\right|}\text{\ensuremath{\left(-\frac{-1}{\sqrt{\left|g\right|}}\frac{\delta S}{\delta e_{\mu}^{*a}}\delta e_{\mu}^{*a}+\frac{1}{\sqrt{\left|g\right|}}\frac{\delta S}{\delta\omega_{ab\mu}}\delta\omega_{ab\mu}\right)}}\\
 & = & \int d^{d}x\sqrt{\left|g\right|}[\left(\nabla_{\mu}+{T^{\rho}}_{\mu\rho}\right)\tau_{\left(c\right)a}^{\mu}\xi^{a}\\
 &&+\left(\tau_{\left(c\right)a}^{\mu}{T^{a}}_{\mu\nu}\xi^{\nu}-S^{\mu ab}\Omega_{ab\mu}\xi^{\nu}\right)],
\end{eqnarray*}
where $\tau_{\left(c\right)a}^{\mu}$ is the canonical energy-momentum
tensor, i.e.,
\[
\tau_{\left(c\right)a}^{\mu}\equiv\frac{-1}{\sqrt{\left|g\right|}}\frac{\delta S}{\delta e_{\mu}^{*a}},
\]
and $S^{\mu ab}$ is the spin current, i.e.,
\[
S^{\mu ab}=\frac{1}{\sqrt{\left|g\right|}}\frac{\delta S}{\delta\omega_{ab\mu}}.
\]
Hence, we have obtained
\begin{equation}
\left(\nabla_{\mu}+{T^{\rho}}_{\mu\rho}\right)\tau_{\left(c\right)a}^{\mu}=\left(\tau_{\left(c\right)a}^{\nu}{T^{a}}_{\mu\nu}-S^{\nu ab}\Omega_{ab\mu\nu}\right)e_{a}^{\mu},
\end{equation}
and similarly, for the chiral energy-momentum tensor $\tau_{\left(c\right)a}^{5\mu}$
of the massless Dirac fermions,
\[
\left(\nabla_{\mu}+{T^{\rho}}_{\mu\rho}\right)\tau_{\left(c\right)a}^{5\mu}=\left(\tau_{\left(c\right)a}^{5\nu}{T^{a}}_{\mu\nu}-S^{\nu ab}\Omega_{ab\mu\nu}\right)e_{a}^{\mu}.
\]
Note that for massive Dirac fermions, the mass term would break the
chiral symmetry and thus lead to an extra mass term in the equation
above.

\subsection{Chiral energy-momentum anomalies}

After deriving the classical equation for the chiral energy-momentum tensor, we are ready to explore the corresponding quantum
anomalies. For the chiral covariant diffeomorphism, i.e.,
\[
\delta\psi=\xi^{\mu}\gamma^{5}D_{\mu}\psi,
\]
one can straightforwardly obtain the Jacobian, i.e.,  $\mathcal{D}\psi\mathcal{D}\bar{\psi}\rightarrow J\left(\xi\right)\mathcal{D}\psi\mathcal{D}\bar{\psi}$
and
\[
\ln J\left(\xi\right)=2\text{Tr}\left(\overleftarrow{D}_{a}\gamma_{5}\right)\xi^{a}.
\]
Similar to the chiral anomaly, the regularized Jacobian is

\begin{eqnarray}
 &  & 2\text{Tr}\left(\overleftarrow{D}_{a}\gamma^{5}e^{\Lambda^{-2}\slashed{\mathcal{D}}\slashed{\mathcal{D}}}\right)\nonumber \\
 & \rightarrow & -2i(-1)\sum_{m\in Z}\frac{1}{m!}\int\int\frac{d^{d}k}{\left(2\pi\right)^{d}}e^{-tk^{2}}k_{a}\left(-2\Lambda^{-2}{T^{d}}k_{d}+2\Lambda^{-2}F\right)^{m},\nonumber
\end{eqnarray}
where $\rightarrow$ means that we have replaced ${T^{d}}_{ab}\left(-i\sigma^{ab}\right)$
by $\frac{1}{2}{T^{d}}_{ab}e^{a}\wedge e^{b}$ with similar notation $F$.
In addition, we have kept only the most divergent term and set the
spin connection to zero. Like in the chiral anomaly case, one can
find that in the zero-curvature limit, the most divergent terms are

\begin{equation}
2\text{Tr}\left(\overleftarrow{D}_{a}\gamma_{5}\right)=(-i)\frac{2\Lambda^{d}}{\left(4\pi\right)^{d/2}}\delta_{ab}\int T^{b}e^{4\Lambda^{-2}\left(T^{e}T^{f}\delta_{ef}\right)}e^{2\Lambda^{-2}F},\label{eq:anomaly_polynomials-1}
\end{equation}
where the spacetime dimension is $2n$. $T^{e}T^{f}\delta_{ef}$ in the exponent reminds us of the Nieh-Yan
term, i.e., $T^{a}\wedge T^{b}\eta_{ab}-e^{*a}\wedge e^{*b}\wedge\Omega_{ab}$,
so in the presence of curvature, it is tempting to conjecture that
\[
2\text{Tr}\left(\overleftarrow{D}_{a}\gamma_{5}\right)=(-i)\frac{2\Lambda^{2n}}{\left(4\pi\right)^{n}}\delta_{ab}T^{b}e^{4\Lambda^{-2}\left(T^{a}\wedge T_{b}\delta_{ab}-e^{*a}\wedge e^{*b}\wedge\Omega_{ab}\right)}e^{2\Lambda^{-2}F}\hat{A}\left(\Omega\right)\dots,
\]
where $\hat{A}\left(\Omega\right)$ is the Dirac genus.

Then, we have found that in two dimensions, 
\[
\nabla_{\mu}\tau_{\left(c\right)a}^{\mu}=\frac{\Lambda^{2}}{4\pi}\frac{\epsilon^{\mu\nu}}{\sqrt{\left|g\right|}}\eta_{ab}{T^{b}}_{\mu\nu}+\dots,
\]
and in $\left(3+1\right)$ dimensions,
\[
\nabla_{\mu}\tau_{\left(c\right)a}^{\mu}=\Lambda^{2}\eta_{ab}\frac{\epsilon^{\mu\nu\rho\sigma}}{\sqrt{|g|}}\frac{T^{b}_{\mu\nu}}{2\pi}\frac{F_{\rho\sigma}}{2\pi}+\dots,
\]
where  $\dots$ means that we have kept only the most divergent 
terms caused by the torsion and the metric is $\eta_{ab}=\text{diag}(1,\ -1,\dots)$. 

\bibliographystyle{apsrev4-1}

\end{document}